\documentclass[11pt,a4paper]{article}
\usepackage{amsmath,amssymb,amsthm}

\def\baselinestretch{1.15}

\newtheorem{teor}{Theorem}
\newtheorem{prop}{Proposition}

\newtheorem{lem}{Lemma}
\newtheorem{definition}{Definition}
\newtheorem*{state*}{Statement}
\newtheorem*{assum*}{Assumption}

\def\beq{\begin{equation}}
\def\eeq{\end{equation}}
\def\bea{\begin{eqnarray}}
\def\eea{\end{eqnarray}}
\def\beann{\begin{eqnarray*}}
\def\eeann{\end{eqnarray*}}
\def\beasn{\begin{sneqnarray}}
\def\eeasn{\end{sneqnarray}}
\def\ben{\begin{enumerate}}
\def\een{\end{enumerate}}
\def\bit{\begin{itemize}}
\def\eit{\end{itemize}}
\def\proof{ ({\sl Proof\/}) }

\def\derpar#1#2{\displaystyle\frac{\partial{#1}}{\partial{#2}}}

\def\moment#1#2#3{{#1}_{#2}, \ldots, {#1}_{#3}}

\def\qed{\ifvmode\Realemovelastskip\fi
{\unskip\nobreak\hfil\penalty50\hbox{}\nobreak\hfil \hbox{\vrule
height1.2ex width1.2ex}\parfillskip=0pt \finalhyphendemerits=0
\par\smallskip}}

\def\vf{\mathfrak{X}}
\def\df{{\mit\Omega}}

\def\d{{\rm d}}

\def\Real{\mathbb{R}}
\def\R{\mathbb{R}}

\def\Tan{{\rm T}}
\def\Lie{\mathop{\mathcal{L}\strut}\nolimits}
\def\inn{\mathop{i\strut}\nolimits}
\def\Cinfty{{\rm C}^\infty}
\def\Ker{\mathop{\rm Ker}}

\def\tabaddress#1{{\small\itshape\begin{tabular}[t]{c}#1
\\[1.2ex]\end{tabular}}}
\def\qed{\ifvmode\removelastskip\fi
{\unskip\nobreak\hfil\penalty50\hbox{}\nobreak\hfil \hbox{\vrule
height1.2ex width1.2ex}\parfillskip=0pt \finalhyphendemerits=0
\par\smallskip}}

\def\relstack#1#2{\mathrel{\mathop{#1}\limits_{#2}}}
\def\weak#1{\relstack{\approx}{#1}}

\newcommand{\ds}{\displaystyle}

\begin{document}

\title{
Constraint algorithm for $k$-presymplectic Hamiltonian systems.
\\
Application to singular field theories
}
\author{%
{Xavier Gr\`acia\thanks{{\itshape email}:
xgracia@ma4.upc.edu}},
{Rub\'en Mart\'{\i}n\thanks{{\itshape email}:
rubenmg@ma4.upc.edu}}
and
{Narciso Rom\'an-Roy\thanks{{\itshape email}:
nrr@ma4.upc.edu}},
\\[1mm]
\tabaddress{Departament de Matem\`atica Aplicada 4\\
Universitat Polit\`ecnica de Catalunya\\
edifici C-3, Campus Nord UPC\\
C/ Jordi Girona 1. E-08034 Barcelona. Spain}}

\date{9 March 2009}

\pagestyle{myheadings}
\markright{\small
X. Gr\`acia, R. Mart\'{\i}n, N. Rom\'an-Roy
--- \sl Constraint algorithm for $k$-presymplectic systems}
{\sffamily
\maketitle
\thispagestyle{empty}

\begin{abstract}
The k-symplectic formulation of field theories is especially simple,
since only tangent and cotangent bundles are needed in its
description.
Its defining elements show a close relationship
with those in the symplectic formulation of mechanics.
It will be shown that
this relationship also stands in the presymplectic case.
In a natural way,
one can mimick the presymplectic constraint algorithm
to obtain a constraint algorithm
that can be applied to $k$-presymplectic field theory,
and more particularly to
the Lagrangian and Hamiltonian formulations of
field theories defined by a singular Lagrangian,
as well as to the unified Lagrangian-Hamiltonian
formalism (Skinner--Rusk formalism)
for $k$-presymplectic field theory.
Two examples of application of the algorithm are also analyzed.

\end{abstract}


\bigskip

\vbox{\raggedleft MSC\,(2000):
35Q99, 53D99, 55R10, 70H45, 70S05
 \\
PACS (1999):
 02.40.Vh, 03.50Kk, 11.10.Ef., 11.10Kk
}\null

}
\clearpage


\section{Introduction}

Systems of singular differential equations have been a matter
of increasing interest in theoretical physics and in some
technical areas such as engineering
of electric networks or control theory.
The fundamental characteristic of these kinds of systems is that
the existence and uniqueness of solutions are not assured.
In particular, this situation arises in mechanics when
dynamical systems described by singular Lagrangians are considered,
and also when considering systems of PDE's associated with field
theories described by singular Lagrangians
(such as, for instance, electromagnetism),
as well as in some other applications related to
optimal control theories.
Furthermore,
these systems do not have a nice Hamiltonian description,
since not all the momenta are available and,
in general, the equations have no solution everywhere.

Bergmann and Dirac were pioneer in solving the problem
for the Hamiltonian formalism of singular mechanical systems,
by developing a constraint
algorithm which gives, in the favourable cases,
a final constraint submanifold
where admissible solutions to the dynamics exist
(in the sense that the dynamical evolution remains on this manifold)
\cite{Dir-64}.
Their main aim was to apply this procedure to field theories.
Afterwards,
a lot of work was done in order to geometrize this algorithm.
The first important step was the work by Gotay {\it et al}
\cite{GNH-78}, and its application to the Lagrangian formalism
\cite{GN-79,GN-80}.
Other algorithms were given later, in order to find
consistent solutions of the dynamical equations
in the Lagrangian formalism of singular systems
(including the problem of finding holonomic solutions)
\cite{BGPR-86,Ka-82,MR-92}, and
afterwards, new geometric algorithms
were developed to be applied both in the Hamiltonian and
the Lagrangian formalisms, as well as to other kinds of more
general systems og singular differential equations
\cite{GP-92,HRT-76,KU-96,MMT-97,Mu-89,RR-94}.

The Lagrangian and Hamiltonian descriptions of field theories is
the natural extension of
time-dependent mechanics. Therefore, in order to understand
the constraint algorithm for field theories
in a covariant formalism, the first step was
to develop the algorithmic procedures for time-dependent systems.
This work was provided mainly in
\cite{CF-93,CLM-94,GM-2005,
LMM-96b,LMMMR-2002,LMM-96c,MS-98,MS-00b,MassVig-00,Vig-00}.

There are several alternative models for describing geometrically
first-order classical field theories.
From a conceptual point of view, the simplest one is the
{\sl $k$-symplectic formalism}, which is the generalization to field
theories of the standard symplectic formalism used as
the geometric framework for describing autonomous dynamical systems.
In this sense, the $k$-symplectic formalism is used to
give a geometric description of certain kinds of field theories: in
a local description, those theories whose Lagrangians or Hamiltonians
depend  on the fields and on the partial derivatives of the
fields, or the corresponding moments, but not on
the space-time coordinates \cite{fam}.
The foundations of the $k$-symplectic formalism are
the $k$-symplectic manifolds \cite{aw,aw3,mt2}.
Historically, it is based on the so-called {\sl polysymplectic formalism}
developed by G\"{u}nther \cite{gun}, who introduced the concept of
{\sl polysymplectic manifold}.
Then, $k$-symplectic manifolds are
polysymplectic manifolds which have {\sl Darboux-type coordinates}
\cite{mt2}.
A natural extension of this formalism is the
{\sl $k$-cosymplectic formalism}, where $k$-cosymplectic manifolds are used
to describe geometrically field theories involving space-time coordinates
or analogous ones, on the Lagrangian or the Hamiltonian
\cite{mod1,mod2}.
This is the generalization
to field theories of the cosymplectic formalism
geometrically describing non-autonomous mechanical systems.
One of the advantages
of these formalisms is that one only needs the tangent and cotangent
bundle of a manifold to develop them.

It is worth noting that G. Sardanashvily \emph{et al}
\cite{Sarda2,Sd-95}
developed a polysymplectic formalism for classical field theories
which differs from the one proposed by G\"unther.
(See also \cite{Kana} for more details on the polysymplectic formalism.)
In addition, we must remark that the soldering form on the linear
frames bundles is a polysymplectic form, and its study and
applications to field theory constitute the $n$-symplectic
formalism developped by L.\,K. Norris
\cite{McN,No2}.

Working within the framework of the $k$-symplectic description for
these theories, we present in this paper a geometric algorithm for
finding the maximal submanifold where there are consistent
solutions to the field equations of singular theories.
This algorithm is a generalization of the
{\sl presymplectic constraint algorithm} for
presymplectic dynamical systems \cite{GNH-78},
and gives an intrinsic description of all the constraint
submanifolds.
The problem is stated in a generic way
for $k$-presymplectic Hamiltonian systems,
in order to give a solution to both
Lagrangian and Hamiltonian field theories, as well as other
possible kinds of systems of partial differential equations. In
this framework, the solutions to these equations are given
geometrically by integrable $k$-vector fields in the manifold
where the equations are stated.
In this way, a constraint algorithm can be developed giving a
sequence of submanifolds which,
in the best case, ends in some final constraint
submanifold where field equations have consistent solutions
($k$-vector fields), although not necessarily integrable.
The general problem of integrability is not addressed in this paper,
only discussed in the examples.
Finally, Lagrangian and Hamiltonian field theories are
particular cases where the above results are applied straightforwardly,
although in the Lagrangian case the problem of finding holonomic
solutions must be also analized.
In addition, the  unified Lagrangian-Hamiltonian
formalism of Skinner--Rusk \cite{skinner2},
which was adapted recently for $k$-symplectic field theories \cite{RRS},
constitutes a framework where this algorithm is applied in
a very natural way.
A description of constraint algorithms
for other geometrical models of field theories
({\sl multisymplectic\/})
was made in \cite{EMR-96,EMR-99b,LMM-96,LMMMR-2005}.

The paper is organized as follows.
In section~2 $k$-symplectic structures are reviewed,
as well as the corresponding Hamiltonian systems.
Section~3 is devoted to $k$-presymplectic Hamiltonian systems
and the presymplectic constraint algorithm.
In section~4 the particular case of field theories described
by a Lagrangian function is considered,
either in Lagrangian or in Hamiltonian formalism.
Finally, the application to the Skinner--Rusk formalism
and two examples are studied in section~5.

Manifolds and maps are assumed to be smooth.
Sum over crossed repeated indices is understood.

\section{$k$-symplectic Hamiltonian systems}

\subsection{$k$-symplectic manifolds. The bundle of $k^1$-covelocities}

\begin{definition}
\label{defaw}
A \emph{$k$-symplectic structure} on a differentiable manifold $M$
of dimension $N=n+kn$
is a family $(\omega^1,\ldots,\omega^k;V)$,
where each $\omega^A$ is a closed $2$-form,
and $V$ is an integrable $nk$-dimensional tangent distribution on $M$
such that
\begin{eqnarray*}
(i) && \omega^A \vert_{V\times V} =0 \ \hbox{for each $A$} ,
\\
(ii) && \cap_{A=1}^{k} \Ker\omega^A = \{0\} .
\end{eqnarray*}
Then $(M,\omega^A,V)$ is called a \emph{$k$-symplectic manifold}.
\end{definition}

\begin{teor}
\cite{aw,mt2}.
Let $(\omega^A,V)$ be a $k$-symplectic structure on~$M$.
For every point of $M$ there exists a neighbourhood $U$
and local coordinates
$(q^i , p^A_i)$ ($1\leq i\leq n$, $1\leq A\leq k$)
such that, on~$U$,
$$
\omega^A=  dq^i\wedge dp^A_i ,
\quad
V =
\left\langle \frac{\partial}{\partial p^1_i}, \dots,
\frac{\partial}{\partial p^k_i} \right\rangle_{i=1,\ldots,n} .
$$
\end{teor}
These are called \emph{Darboux} or \emph{canonical coordinates}
of the $k$-symplectic manifold.

The canonical model of a $k$-symplectic manifold is
$(\Tan^1_k)^*Q= \Tan^*Q\oplus \stackrel{k}{\dots} \oplus \Tan^*Q$,
the {\sl bundle of $k^1$-covelocities} of
an $n$-dimensional differentiable manifold $Q$,
which has the natural projections
$$
\begin{array}{ccccccc}
\pi^A \colon (\Tan^1_k)^*Q & \rightarrow & \Tan^*Q
& ; &
\pi^1_Q \colon (\Tan^1_k)^*Q & \to & Q
\\
(q;\alpha^1_q, \ldots ,\alpha^k_q) &\mapsto& (q;\alpha^A_q)
& ; &
(q;\alpha^1_q, \ldots ,\alpha^k_q)&\mapsto& q
\end{array} \ .
$$
$(\Tan^1_k)^*Q$ is endowed with the canonical forms
$$
\theta^A=(\pi^A)^*\theta ,
\quad
\omega^A=(\pi^A)^*\omega=-(\pi^A)^*\d\theta=-\d\theta^A ,
$$
where $\theta$ and $\omega$ are the Liouville $1$-form and
the canonical symplectic form on $\Tan^*Q$.

If $(q^i)$ ($1\leq i\leq n$) are local coordinates on $U \subset Q$,
the induced coordinates $(q^i ,p^A_i)$ ($1\leq A\leq k$) on
$(\pi^1_Q)^{-1}(U)$ are given by
$$
q^i(q;\alpha^1_q, \ldots ,\alpha^k_q)= q^i(q) ,
\quad
p^A_i(q;\alpha^1_q, \ldots ,\alpha^k_q)=
\alpha^A_q\left(\frac{\partial}{\partial q^i}\Big\vert_q\right) .
$$
Then we have
$$
\theta^A = p^A_i\d q^i ,
\quad
\omega^A =
\d q^i\wedge\d p^A_i .
$$
Thus, the triple  $((\Tan^1_k)^*Q,\omega^A, V)$,
where $V=\Ker \Tan \pi^1_Q$,
is a $k$-symplectic manifold,
and the natural coordinates in $(\Tan^1_k)^*Q$ are Darboux coordinates.

\subsection{k-vector fields and integral sections}

Let
$\Tan^{1}_{k}M = \Tan M \oplus \stackrel{k}{\dots} \oplus \Tan M$
be the {\sl bundle of $k^1$-velocities} of a differentiable manifold~$M$.
It is endowed with the natural projections
\beq
\begin{array}{ccccccc}
\tau^A\colon \Tan^1_kM &\rightarrow& \Tan M
& ; &
\tau^1_M\colon \Tan^1_kM &\to& M
\\
(q,{v_1}_q,\ldots ,{v_k}_q) &\mapsto& (q;{v_A}_q)
& ; &
(q,{v_1}_q,\ldots ,{v_k}_q)&\mapsto& q
\end{array}
\label{natpro}
\eeq

\begin{definition}
\label{kvector}
A {\em $k$-vector field} on a manifold $M$ is a section
${\bf X} \colon M \longrightarrow \Tan^1_kM$ of the projection~$\tau_M^1$.
\end{definition}

Therefore, giving a $k$-vector field ${\bf X}$ is the same as
giving $k$ vector fields $X_{1}, \dots, X_{k}$ on $M$,
obtained as
$X_A=\tau_A\circ{\bf X}$.
We denote ${\bf X}=(X_1, \ldots, X_k)$.

\paragraph{Remark}
The term $k$-vector field on~$M$ is more often applied
to the sections of the bundle $\Lambda^k\Tan M\to M$,
that is, contravariant skew-symmetric tensor fields of order~$k$.
The $k$-vector fields ${\bf X}=(X_1, \ldots, X_k)$ used here
lead to a particular class of such tensor fields,
the {\sl decomposable} ones,
$X_1 \wedge \ldots \wedge X_k$,
which can be associated with distributions on~$M$.

\begin{definition}
\label{integsect}
An {\rm integral section} of the $k$-vector field
${\bf X}=(X_{1},\dots, X_{k})$
is a map
$\phi \colon J \rightarrow M$,
defined on an open set $J \subset \Real^k$,
such that
$$
\Tan \phi \circ \frac{\partial}{\partial t^A}
=
X_{A} \circ \phi ,
$$
where $t=(t^1,\ldots,t^k)$ denote the canonical coordinates
of~$\Real^k$.
\\
Equivalently,
an integral section satisfies the equation
$$
\phi^{(1)} = {\bf X} \circ \phi ,
$$
where $\phi^{(1)} \colon J \to \Tan^1_kM$
is the {\rm first prolongation} of $\phi$ to $\Tan^1_kM$
defined by
$$
\phi^{(1)}(t)
=
\left(
\phi(t),
\Tan \phi \left(\frac{\partial}{\partial t^1}\Big\vert_{t}\right),
\ldots,
\Tan \phi \left(\frac{\partial}{\partial t^k}\Big\vert_{t}\right)
\right)\ .
$$
A $k$-vector field ${\bf X}$ is {\rm integrable} if
every point of~$M$ belongs to the image of an integral section
of~${\bf X}$.
\end{definition}

In coordinates, write
$\ds X_A = X_A^i \frac{\partial}{\partial x^i}$.
The $\phi$ is an integral section of $\mathbf{X}$ if, and only if,
the following system of partial differential equations holds:
$$
\frac{\partial \phi^i}{\partial t^A} = X_A^i(\phi) \ .
$$

\begin{prop}
A $k$-vector field ${\bf X}=(X_1, \ldots, X_k)$ is integrable
if, and only if,
$[X_A,X_B] = 0$
for each $A,B$.
\end{prop}
This is the geometric expression
of the integrability condition of the preceding differential equation
(see, for instance, \cite{Lee} or \cite{Die}).

\subsection{Hamiltonian systems}

\begin{definition}
Let $(M,\omega^A,V)$
be a $k$-symplectic manifold,
and $\alpha \in \df^1(M)$ a closed form.
$(M,\omega^A,V,\alpha)$ is said to be a
{\rm $k$-symplectic Hamiltonian system}.
\end{definition}

As $\alpha$ is closed, for every point of $M$ there exists a
neighbourhood $U\subset M$ and a function $H \in \Cinfty(U)$
such that $\alpha=\d H$ on $U$.
This function is called a {\sl local Hamiltonian function}.
If $\alpha$ is exact, then
$H\in\Cinfty(M)$ is called  a {\sl (global) Hamiltonian function}.
These functions are unique up to a constant on each connected component of~$M$.
From now on, we will write $\alpha=\d H$.

The
\emph{Hamilton--de Donder--Weyl (HDW) equation}
for a map
$\psi \colon J \to M$
($J \subset \R^k$)
is
\begin{equation}
 \inn({\psi^{(1)}_A}) \omega^A = \d H\circ\psi\ .
\label{he20}
\end{equation}
In canonical coordinates this reads
$$
\frac{\partial\psi^i}{\partial t^A} =
\frac{\partial H}{\partial p^A_i} ,
\qquad
\frac{\partial\psi^A_i}{\partial t^A} =
- \frac{\partial H}{\partial q^i} .
$$
where
$\psi = (\psi^i,\psi^A_i)$.
Recall that, according to our conventions,
a sum $\sum_A$ is understood whenever the index~$A$ appears
twice in upper and lower position.

\smallskip

In order to give an alternative geometrical interpretation
of these equations,
we introduce the set $\vf^k_H(M)$ of those $k$-vector fields
${\bf X}=(X_1,\dots,X_k)$ on $M$
which are solutions of the {\sl geometric field equation}
\begin{equation}
\label{generic0}
\inn(X_A)\,\omega^A = \d H .
\end{equation}

For $k$-symplectic Hamiltonian systems,
solutions of equation~(\ref{generic0}) always exist
(this is a consequence of
the lemma and the theorem in the next section).
They are neither unique, nor necessarily integrable.

In canonical coordinates of $M$, writing
$\ds
X_A =
(X_A)^i\frac{\partial}{\partial q^i}+
(X_A)_i^B\frac{\partial}{\partial p_i^B}
$,
equation (\ref{generic0}) reads
$$
\ds
\frac{\partial H}{\partial q^i} =  -(X_A)^A_i \,,
\quad
\frac{\partial H}{\partial p^A_i} = (X_A)^i .
$$

This geometric field equation for ${\bf X}$ is an alternative formulation
of the HDW equation in the following sense:
\begin{prop}
\label{prop1}
Let ${\bf X}=(X_1,\dots,X_k)$ be an integrable $k$-vector field in $M$.
Every integral section $\psi \colon J \to M$ of ${\bf X}$
satisfies the HDW equation (\ref{he20})
if, and only if,
${\bf X} \in \vf^k_H(M)$.
\end{prop}

Note however that equations (\ref{he20}) and (\ref{generic0})
cannot, in general, be considered as fully equivalent:
a solution to the HDW equations may not be
an integral section of some integrable $k$-vector field on~$M$.
Solutions $\psi$ that are integral sections of some
${\bf X} \in \vf^k_H(M)$
will be called {\em admissible},
and we will restrict our attention to them.

\section{$k$-presymplectic system and constraint algorithms}

\subsection{$k$-presymplectic Hamiltonian systems}

To consider singular field theories we have to drop some assumptions
in the definition of a $k$-symplectic structure.
So, a family $(\omega^1,\ldots,\omega^k)$
ok $k$ closed 2-forms on a smooth manifold $M$
will be called a {\em $k$-presymplectic structure};
accordingly, $(M,\omega^A)$ will be called a
{\em $k$-presymplectic manifold}.

The simplest example of a $k$-presymplectic manifold
is provided by any submanifold of a $k$-symplectic manifold:
the pull-back of the $k$ 2-forms by the inclusion map
yields $k$ 2-forms on the submanifold.

In some particular $k$-presymplectic manifolds
one can find Poisson-like coordinates,
but it is an open question to characterize the necessary
and sufficient conditions for these coordinates to exist.

Given a closed 1-form $\alpha \in \df^1(M)$,
$(M,\omega^A,\alpha)$ is said to be a
{\em $k$-presymplectic Hamiltonian system}.
As above, we will write $\alpha=\d H$ (locally or globally).

Then we can also consider the
{Hamilton--de Donder--Weyl equation}
$
\inn({\psi^{(1)}_A}) \omega^A = \d H\circ\psi
$,
and also the set
$\vf^k_H(M)$
of $k$-vector fields ${\bf X}$
that satisfy the geometric field equation
$
\inn(X_A)\,\omega^A = \d H
$.

For $k$-presymplectic systems the existence of solutions
of this equation is not assured everywhere on $M$.
We will analyze the existence of solutions on a certain
submanifold of $M$.

\subsection{Statement of the problem}

The problem we wish to solve arises from the Lagrangian and
Hamiltonian $k$-presymplectic formalisms in field theories,
although other kinds of systems could also be stated in this way.

\begin{state*}
Let $(M,\omega^A,\d H)$ be a $k$-presymplectic Hamiltonian system.
We want to find a submanifold $C$ of $M$
and integrable $k$-vector fields ${\bf X}=(X_1,\ldots,X_k)\in\vf^k(M)$
such that
\beq
\label{E}
\inn(X_{A}) \omega^A \weak{C}
\d H
\eeq
(this means equality on the points of~$C$)
and
$$
\hbox{${\bf X}$ is tangent to $C$}
$$
(this means that $X_1,\ldots,X_k$ are tangent to $C$).
\end{state*}

As stated in the introduction,
we will focus on the consistency of the equation
and will not address the integrability condition in generality.

Given a submanifold $C$ of $M$,
with natural embedding $\jmath_C \colon C \hookrightarrow M$,
let $\Tan^k_1\jmath_C \colon \Tan^1_kC \to \Tan^1_kM$ be
the natural extension of $\jmath_C$ to the $k$-tangent bundles,
and denote its image as
$\underline{\Tan^1_kC} = \Tan^k_1\jmath_C(\Tan^1_kC)$.

We can define the map
\beq
\begin{array}{rccc}
\flat_k \colon & \Tan^1_kM & \to & \Tan^*M
\\
   & (p;v_{p_1},\ldots,v_{p_k}) & \mapsto & (p,\inn(v_{A_p})\omega^A_p)
\end{array}
\label{fdtal}
\eeq
and denote by $(\Tan C)^\perp_{\flat_k}$ the annihilator
of the image of $\underline{\Tan^1_kC}$ by $\flat_k$; that is,
$$
(\Tan C)^\perp_{\flat_k}
=
[\flat_k(\underline{\Tan^1_kC})]^0
=
\{ u_p \in \Tan M \mid
\forall (v_{p1},\ldots,v_{pk}) \in \underline{T^1_kC} , \
\langle {\small \sum} \inn(v_{pA}) \omega^A_p , u_p \rangle =0 \} .
$$
We call $(\Tan C)^\perp_{\flat_k}$ the
{\sl $k$-presymplectic orthogonal complement} of
$\underline{\Tan^1_kC}$
in $\Tan^1_kM$.

In particular, for $C=M$ we have:

\begin{lem}
$\ds
(\Tan M)^\perp_{\flat_k}
=
\{ (p,u_p) \in \Tan M \mid u_p \in \bigcap_{A=1}^k \Ker\omega^A_p\}
$
\end{lem}
\proof\
For every $p\in M$ and $(v_{p1},\ldots,v_{pk})\in(\Tan^1_k)_pM$,
if $u_p \in \bigcap_{A=1}^k \Ker\omega^A_p$,
then we have
$(\inn (v_{Ap})\omega^A_p)(u_p) =
-\inn(v_{Ap})\inn(u_p)\omega^A_p =
0$,
and therefore $u_p\in(\Tan_pM)^\perp_{\flat_k}$.

Conversely, if $u_p\in(\Tan_pM)^\perp_{\flat_k}$,
then $(\inn(v_{Ap})\omega^A_p)(u_p)=0$
for every $(v_{p1},\ldots,v_{pk})\in(\Tan^1_k)_pM$.
Then taking any $(v_{p1},0,\ldots,0)$ with $v_{p1} \neq 0$
we conclude that $u_p \in \Ker \omega^1_p$;
and analogously for the others.
\qed

The main result is the following:
\begin{teor}
\label{main}
Let $C$ be a submanifold of~$M$.
The following conditions are equivalent:
\bit
\item
there exists a $k$-vector field ${\bf X}=(X_1,\ldots,X_k)\in\vf^k(M)$,
tangent to $C$, such that
equation (\ref{E}) holds
\vadjust{\kern -5mm}
\item
\null\ \vadjust{\kern -5mm}
\beq
\label{nsc}
\inn(Y_p)(\d H)_p=0 \quad
\mbox{\rm for every $p\in C$, $Y_p\in (\Tan_pC)^\perp_{\flat_k}$.}
\eeq
\eit
\end{teor}
\proof
($\Longrightarrow$)
If there exists a $k$-vector field ${\bf X}=(X_1,\ldots,X_k)\in\vf^k(M)$,
tangent to $C$ such that equation (\ref{E}) holds, then,
for every $p\in C$ and $Y_p\in (\Tan_pC)^\perp_{\flat_k}$,
$$
0 =
[\inn(X_{Ap})\omega^A_p](Y_p) =
\inn(Y_p)\inn(X_{Ap})\omega^A_p=\inn(Y_p)(\d H)_p \,.
$$

($\Longleftarrow$)
If (\ref{nsc}) holds, then
$$
(\d H)_p\in[(\Tan C)^\perp_{\flat_k}]^0=
[[\flat_k(\underline{\Tan^1_kC})]^0]^0=\flat_k(\underline{\Tan^1_kC}) ,
$$
and hence there exists
$(X_{p1},\ldots,X_{pk}) \in \underline{(\Tan^1_k)_pC}$
such that (\ref{E}) holds.
\qed

\subsection{$k$-presymplectic constraint algorithm}
\protect\label{precoal}

The application of the above result leads to an algorithmic procedure
which gives a sequence of  subsets
$\ldots \subset C_j \subset \ldots C_2\subset C_1\subset M$.
We will assume that:

\begin{assum*}
Every subset $C_j$ of this sequence is a regular submanifold of $M$.
\end{assum*}

These submanifolds are sequentially obtained from the analysis
of the consistency of a linear equation, namely eq.~(\ref{E})
at each point:
$$
\inn(X_{Ap}) \omega^A_p = (\d H)_p .
$$

First, $C_1 \hookrightarrow M$ is
the submanifold of~$M$ where this equation is consistent:
$$
C_1 =
\{ p \in M \mid
\exists {\bf X}_p
\ \mbox{\rm such that}\
\inn(X_{Ap}) \omega^A_p = (\d H)_p \} .
$$
So, there exist $k$-vector fields ${\bf X}$ on~$M$
which satisfy equation~(\ref{E}) on the sumbanifold~$C_1$.
However, in general these ${\bf X}$ may not be tangent to $C_1$.
Therefore, we consider the submanifold
$$
C_2 =
\{ p \in C_1 \mid
\exists {\bf X}_p \in \Tan^1_k(C_1)
\ \mbox{\rm such that}\
\inn(X_{Ap}) \omega^A_p = (\d H)_p \}
\,,
$$
and so on.
Following this process, we obtain a sequence of constraint submanifolds
$$
\ldots \hookrightarrow C_j \hookrightarrow \ldots C_2\hookrightarrow C_1
\hookrightarrow M
$$
where, taking into account Theorem~\ref{main},
each submanifold $C_j$ is geometrically defined by
$$
C_j=
\{ p\in C_{j-1} \mid \inn(Y_p) (\d H)_p = 0
\ \mbox{\rm for every $Y_p \in (\Tan_p C_{j-1})^\perp_{\flat_k}$}
\}.
$$
For every $j \geq 1$, $C_j$ is called the
{\sl $j$th constraint submanifold}.

If we denote by
$\vf(C_j)^\perp_{\flat_k}$
the set of vector fields $Y$ in $M$ such that
$Y_p \in (\Tan_p C_j)^\perp_{\flat_k}$,
then one can obtain constraint functions $\{\xi_\mu\}$ defining each $C_j$
from a local basis
$\{ Z_1,\ldots,Z_r\}$ of vector fields of
$\vf(C_{j-1})^\perp_{\flat_k}$
by setting
$\xi_\mu = \inn(Z_\mu) \d H$.

The technical procedure to obtain these constraints is
the following:
\begin{itemize}
\item
To obtain a local basis
$\{ Z_1,\ldots,Z_r\}$ of vector fields of $\bigcap_{A=1}^k \Ker\omega^A$.
\item
To apply Theorem~\ref{main}
to obtain a set of independent constraint functions
$\xi_\mu=\inn(Z_\mu)\d H$, defining $C_1\hookrightarrow M$.
\item
To calculate ${\bf X}=(X_1,\ldots,X_k)$, solutions to (\ref{E}) on $C_1$.
\item
To impose the tangency condition of $X_1,\ldots,X_k$
 on the constraints $\xi_\mu$.
\item
To iterate the last item until no new constraints appear.
\end{itemize}

This is the {\sl $k$-presymplectic constraint algorithm}.
We have two possibilities:
\bit
\item
There exists an integer $j>0$ such that $C_{j+1}=C_j\equiv C_f$.
In this case, $C_f$ is called the {\sl final constraint submanifold},
and there exist a family of $k$-vector fields
${\bf X}^f=(X^f_1,\ldots,X^f_k)$ in $M$,
tangent to $C_{f}$, such that (\ref{generic0}) holds on $C_{f}$,
that is,
\beq
[\inn( X^f_A)\omega^A-\d H]\vert_{C_f} = 0 .
\label{form1}
\eeq
This is the situation which is interesting to us.
\item
There exists an integer $j>0$ such that $C_j = \emptyset$.
This means that the equations have no solution
on a submanifold of $M$.
\eit


\section{$k$-symplectic field theory}

\subsection{The bundle of $k^1$-velocities}

The Lagrangian formalism of $k$-symplectic field theories uses
the bundle of $k^1$-velocities of a manifold as phase space.
First we introduce the canonical structures which this manifold
is endowed with.

Let $\Tan^1_kQ=\Tan Q\oplus \stackrel{k}{\dots} \oplus \Tan Q$
be the {\sl bundle of $k^1$-velocities} of $Q$,
with natural projections
$\tau^A\colon \Tan^1_kQ \rightarrow \Tan Q$ and
$\tau^1_Q\colon \Tan^1_kQ \to Q$,
given in (\ref{natpro}).

If  $(q^i)$ are local coordinates on $U \subset Q$,
the induced coordinates $(q^i ,v_A^i)$ on $(\tau^1_Q)^{-1}(U)$ are
$$
q^i(v_{1q},\ldots,v_{kq}) = q^i(q)
\quad ,\quad
v_A^i(v_{1q},\ldots,v_{kq}) = v_{Aq}(q^i) \, .
$$

\medskip

For $Z_q\in \Tan_qQ$,  its {\sl vertical $A$-lift} at
$(v_{1q},\ldots,v_{kq}) \in \Tan_k^1Q$ is the vector $(Z_q)^{V_A}$,
tangent to the
fiber $(\tau^1_Q)^{-1}(q)\subset \Tan_k^1Q$, given by
$$
(Z_q)^{V_A} (v_{1q},\ldots,v_{kq}) =
\frac{d}{ds}\Big|_{s=0}
\!\!\!\!\!\!
(v_{1q},\ldots,v_{(A-1)q},v_{Aq}+sZ_q,v_{(A+1)q},\ldots,v_{kq})
$$
If
$\ds
Z_q = a^i \frac{\partial}{\partial q^i} \Big\vert_q$,
then
$\ds
(Z_q)^{V_A}(v_{1q},\ldots,v_{kq})
=
a^i
\frac{\partial}{\partial v^i_A} \Big\vert_{(v_{1q},\ldots,v_{kq})}
$.

\medskip

The {\em canonical $k$-tangent structure} on $T^1_kQ$ is the set
$(S^1,\ldots,S^k)$ of tensor fields  of type $(1,1)$ defined by
$$
S^A(w_q)(Z_{w_q})= (\Tan_{w_q}(\tau^1_Q)(Z_{w_q}))^{V_A}(w_q)\ ,
$$
for
$w_q = (v_{1q},\ldots,v_{kq}) \in T^1_kQ$,
$\ Z_{w_q}\in T_{w_q}(T^1_kQ)$.

In coordinates we have
$S^A = \frac{\partial}{\partial v^i_A} \otimes \d q^i$.

The {\sl Liouville vector field}
$\Delta\in\vf(\Tan^1_kQ)$  is the infinitesimal generator of the flow
$\psi\colon\Real\times \Tan^1_kQ\longrightarrow \Tan^1_kQ$
$$
\psi(s;v_{1q},\ldots,v_{kq}) =
(e^s v_{1q},\ldots,e^s v_{kq}).
$$
Observe that
$\Delta=\Delta_1+\ldots+\Delta_k$,
where each $\Delta_A\in\vf(\Tan^1_KQ)$
is the infinitesimal generator of the flow
$\psi^A\colon \R \times \Tan^1_kQ \longrightarrow \Tan^1_kQ$
$$
\psi^A(s;v_{1q},\ldots,v_{kq})=
(v_{1q},\ldots,v_{(A-1)q}, e^s v_{Aq},v_{(A+1)q},\ldots,v_{kq})
$$
In local coordinates we have \
$\Delta = \sum_{A=1}^k \Delta_A=v^i_A \derpar{}{v_A^i}$.

\medskip

Now we want to characterize the integrable
$k$-vector fields on $T^1_kQ$ such that their integral sections
are first prolongations $\phi^{(1)}$ of maps $\phi\colon\Real^k\to Q$.
Remember that a $k$-vector field in $T^1_kQ$ is a section
$\mathbf{\Gamma} \colon T^1_kQ \longrightarrow T^1_k(T^1_kQ)$
of the canonical projection
$\tau_{T^1_kQ}\colon T^1_k(T^1_kQ)\to T^1_kQ$.
Then:

\begin{definition}
\label{sode0}
A {\rm second order partial differential equation ({\sc sopde})}
is a $k$-vector field $\mathbf{\Gamma}=(\Gamma_1,\ldots,\Gamma_k)$
on $T^1_kQ$
which is also a section of the projection
$T^1_k\tau \colon T^1_k(T^1_kQ) \rightarrow T^1_kQ$; that is,
$$
T^1_k\tau\circ\mathbf{\Gamma}={\rm Id}_{T^1_kQ} \ ,
$$
or, what is equivalent,
$\Tan_{w_q} \tau \cdot \Gamma_A(w_q) = v_{Aq}$,
for $w_q=(v_{1q},\ldots, v_{kq})\in T^1_kQ$.
\end{definition}

If the local expression of the $k$-vector field
$\mathbf{\Gamma} = (\Gamma_A)$ on $T^1_kQ$
is
$\ds
\Gamma_A =
(\Gamma_A)^i \frac{\partial}{\partial  q^i} +
(\Gamma_A)^i_B \frac{\partial}{\partial v^i_B}
$,
then $\mathbf{\Gamma}$ is a {\sc sopde} iff
$(\Gamma_A)^i = v^i_A$:
$$
\Gamma_A(q^i,v^i_A) =
v^i_A\frac{\partial} {\partial q^i}+
(\Gamma_A)^i_B \frac{\partial} {\partial v^i_B} \, ,
$$
where
$(\Gamma_A)^i_B$
are functions locally defined in $T^1_kQ$.

If $\psi\colon\Real^k \to T^1_kQ$ is an integral section of
$\mathbf{\Gamma}=(\Gamma_1,\ldots,\Gamma_k)$, locally given by
$\psi(t)=(\psi^i(t),\psi^i_B(t))$,
then from the last expression and Definition \ref{integsect}
we deduce
$$
\frac{\partial\psi^i} {\partial t^A}\Big\vert_t=\psi^i_A(t) \, ,
\quad
\frac{\partial\psi^i_B} {\partial t^A}\Big\vert_t=(\Gamma_A)^i_B(\psi(t))
\, .
$$

\begin{prop}
\label{sope1}
Let $\mathbf{\Gamma}=(\Gamma_1,\ldots,\Gamma_k)$ be an integrable
{\sc sopde}.
If $\psi$ is an integral section of ${\bf \Gamma}$
then $\psi=\phi^{(1)}$, where $\phi^{(1)}$ is the first
prolongation of the map
$\phi=\tau\circ\psi \colon
\Real^k \stackrel{\psi}{\to} T^1_kQ \stackrel{\tau}{\to} Q$,
and $\phi$ is a solution of the system of second order partial
differential equations
\beq
\label{nn1}
\frac{\partial^2 \phi^i}{\partial t^A\partial t^B}(t)=
(\Gamma_A)^i_B\left(\phi^i(t),\frac{\partial\phi^i}{\partial
t^C}(t)\right) \, .
\eeq
Conversely, if $\phi\colon\Real^k \to Q$ is any map satisfying
(\ref{nn1}), then $\phi^{(1)}$ is an integral section of
$\mathbf{\Gamma}=(\Gamma_1,\ldots,\Gamma_k)$.
\end{prop}

From (\ref{nn1}) we deduce that if $\mathbf{\Gamma}$ is an
integrable {\sc sopde} then $(\Gamma_A)^i_B=(\Gamma_B)^i_A$.

Finally, using the canonical $k$-tangent structure of $T^1_kQ$,
we have that a $k$-vector field
$\mathbf{\Gamma}=(\Gamma_1,\ldots,\Gamma_k)$ on
$T^1_kQ$ is a {\sc sopde}
if, and only if,
$S^A(\Gamma_A)=\Delta_A$ ($A$ fixed).

\subsection{$k$-symplectic Lagrangian field theory}

Let $L\in\Cinfty(\Tan^1_kQ)$ be a {\sl Lagrangian function}.
We define the {\sl Lagrangian forms}
$$
\theta_L^A = {}^t(S^A) \circ \d L \in \df^1(\Tan^1_kQ)
\ , \quad
\omega_L^A = -\d\theta_L^A \in \df^2(\Tan^1_kQ)\ .
$$
and the {\sl Lagrangian energy function}
$$
E_L = \Delta (L) - L
\in \Cinfty(\Tan^1_kQ)\ .
$$
They have local expressions
$
\ds
\theta_L^A =
\frac{\partial L}{\partial v^i_A}\, \d q^i
$,
$\ds
\omega_L^A =
\d q^i \wedge \d\left(\frac{\partial L}{\partial v^i_A}\right)
$,
$\ds
E_L=v^i_A\frac{\partial L}{\partial v^i_A}-L
$.

We introduce the {\sl Legendre map} of~$L$,
which is its fibre derivative
$FL \colon \Tan^1_kQ \longrightarrow (\Tan^1_k)^*Q$.
It can be defined as follows:
for $q \in Q$, $u_q \in \Tan_qQ$,
$(v_{1q},\dots,v_{kq}) \in (\Tan^1_k)_qQ$,
$$
[FL(v_{1q},\dots,v_{kq})]^A(u_q) =
\frac{d}{ds}
L(v_{1q},\dots,v_{Aq}+su_q,\ldots,v_{k_q}) \vert_{s=0} .
$$
Locally,
$\ds
FL(q^i,v^i_A) =
\left(q^i, \frac{\partial L}{\partial v^i_A}\right)
$.
Furthermore, we have that
$\theta_L^A = FL^*(\theta^A)$,
$\omega_L^A = FL^*(\omega^A)$.

\medskip

\begin{definition}
The Lagrangian $L$ is {\rm regular} if the following equivalent
conditions hold:
\begin{enumerate}
\item
$\ds
\left(\frac{\partial^2 L}{\partial v^i_A \partial v^j_B}\right)
$
is everywhere nonsingular.
\item
The second fibre derivative
$FL \colon
\Tan^1_kQ \longrightarrow (\Tan^1_k)^*Q \otimes (\Tan^1_k)^*Q$
is everywhere nonsingular.
\item
 $FL$ is a local diffeomorphism.
\item
$(\Tan^1_kQ,\omega_L^A,V= \Ker \Tan\tau^1_Q)$
is a $k$-symplectic manifold.
\end{enumerate}
The Lagrangian $L$ is called {\rm hyperregular} if
$FL$ is a global diffeomorphism.
\end{definition}

We must point out that, in field theories, the notion of regularity
is not uniquely defined (for other approaches see, for instance,
\cite{Be-84,De-78,Kr-87,KS-01a,KS-2001}).

Our purpose, however, is the study of singular Lagrangians, {\em i.e.},
those which are not regular.
Following \cite{GN-79}, we will deal with singular Lagrangians
 satisfying some regularity conditions:

\begin{definition}
A singular Lagrangian $L$ is
{\rm almost-regular} if
\begin{enumerate}
\item
$\mathcal{P}:= FL(\Tan^1_kQ)$
is a closed submanifold of $(\Tan^1_k)^*Q$.
\item
$FL$ is a submersion onto its image.
\item
The fibres $FL^{-1}(p)$, for every $p \in \mathcal{P}$,
are connected submanifolds of $\Tan^1_kQ$.
\end{enumerate}
\end{definition}

If $L$ is regular, $(\Tan^1_kQ,\omega^A_L,E_L)$ is a
{\rm $k$-symplectic Lagrangian system},
otherwise it is a {\rm $k$-presymplectic Lagrangian system}.
Therefore,
$(\Tan^1_kQ,\omega_L^A,\d E_L)$ is a
$k$-symplectic or a $k$-presymplectic Hamiltonian system,
depending on the regularity of $L$.

\medskip

In a natural chart of $\Tan^1_kQ$ we have
the {\sl Euler--Lagrange (EL) equations} for $L$, which are
\begin{equation}
\label{ELe}
\frac{\partial}{\partial t^A}
\left(\frac{\displaystyle\partial L}{\partial
v^i_A}\Big\vert_{\varphi(t)}\right)= \frac{\partial L}{\partial
q^i}\Big\vert_{\varphi(t)}
 \  , \quad
v^i_A(\varphi(t))= \frac{\partial\varphi^i}{\partial t^A} \ ,
\end{equation}
whose solutions are maps $\varphi \colon \Real^k \to \Tan^1_kQ$
that are first prolongations to $\Tan^1_kQ$
of maps $\phi=\tau^1_Q\circ\varphi\colon\Real^k \to Q$;
that is, $\varphi$ are {\sl holonomic}.
We will show that these equations can be given a geometric
interpretation using the $k$-presymplectic structure.

Indeed, consider a map
$\varphi \colon \Real^k \to \Tan^1_kQ$
which is holonomic.
Then the Euler--Lagrange equations for $\varphi$
can be also written as
\begin{equation}
\label{ELk}
\inn(\varphi_A^{(1)}) \omega_L^A = \d E_L \, .
\end{equation}

As in our general discussion on $k$-presymplectic Hamiltonian systems,
a convenient way to represent the solutions of these equations
can be set in terms of $k$-vector fields.
Let us introduce the set
$\vf^k_L(\Tan^1_kQ)$ of $k$-vector fields
${\bf \Gamma}=(\Gamma_1,\dots,\Gamma_k)$ in $\Tan^1_kQ$
which are solutions of
\begin{equation}
\label{genericEL}
\inn(\Gamma_A)\omega_L^A=\d E_L\, .
\end{equation}
If
$\ds
\Gamma _A  =
( \Gamma _A)^i \frac{\partial}{\partial  q^i} +
( \Gamma _A)^i_B\frac{\partial}{\partial v^i_B}
$
locally,
then (\ref{genericEL}) is equivalent to
\bea
\label{eqcoor}
\left(
\frac{\partial^2 L}{\partial q^i \partial v^j_A} -
\frac{\partial^2 L}{\partial q^j \partial v^i_A}
\right) \,
(\Gamma_A)^j
-
\frac{\partial^2 L}{\partial v_A^i \partial v^j_B} \,
(\Gamma_A)^j_B
&=&
v_A^j \frac{\partial^2 L}{\partial q^i\partial v^j_A}
-
\frac{\partial L}{\partial q^i}
\nonumber
\\
\frac{\partial^2 L}{\partial v^j_B \partial v^i_A} \, (\Gamma_A)^i
&=&
\frac{\partial^2 L}{\partial v^j_B\partial v^i_A} \, v_A^i
\, .
\eea
If, in addition,
${\bf \Gamma}$ is required to be a {\sc sopde},
{\it i.e.}\ $(\Gamma_A)^i = v_A^i$,
then the above equations are equivalent to
$$
\frac{\partial^2 L}{\partial q^j \partial v^i_A} v^j_A +
\frac{\partial^2 L}{\partial v_A^i\partial v^j_B}(\Gamma_A)^j_B
=
\frac{\partial  L}{\partial q^i}
\, .
$$
These equations imply that,
if ${\bf \Gamma}$ is an integrable {\sc sopde},
its integral sections are holonomic and
they are solutions to the EL-equations.


If $L$ is regular,
solutions to (\ref{genericEL}) always exist, although
they are neither unique, nor necessarily integrable.
However, if ${\bf \Gamma}$ is
integrable, then the second group of equations (\ref{eqcoor})
imply that its integral sections are holonomic and they are solutions to the
EL-equations. Hence ${\bf \Gamma}$ is a {\sc sopde}.

If $L$ is not regular then, in general,
equations (\ref{genericEL}) have no
solutions everywhere in $\Tan^1_kQ$ but, in the most favourable situations,
they do in a submanifold of $\Tan^1_kQ$ which is obtained by applying
the $k$-presymplectic constraint algorithm developed in
Section \ref{precoal}.
Nevertheless, solutions to equations (\ref{genericEL})
are not necessarily {\sc sopde}'s
(unless it is required as an additional condition).
In addition, if they are integrable, their integral sections are not necessarily holonomic,
and thus they are not solutions to the EL-equations (\ref{ELe}).
The geometric analysis of this problem must be done in a separate way.
(For the multisymplectic formalism of field theories,
a study of this problem can be found in \cite{LMMMR-2005}).

\subsection{$k$-symplectic Hamiltonian field theory}

The Hamiltonian formalism of $k$-symplectic regular field theories uses
the bundle of $k^1$-covelocities of a manifold as phase space.

So, consider the  $k$-symplectic manifold $((\Tan^1_k)^*Q,\omega^A,V)$,
and
let $H\in\Cinfty((\Tan^1_k)^*Q)$ be a {\sl Hamiltonian function}.
Then $((\Tan^1_k)^*Q,\omega^A,\d H)$ is a
$k$-symplectic Hamiltonian system.

In particular, if $(\Tan^1_kQ,\omega^A_L,\d E_L)$ is a
{\sl Lagrangian system}, then:
\begin{itemize}
 \item
If $L$ is hyperregular,
we may define the Hamiltonian $H=E_L\circ FL^{-1}$, and
$((\Tan^1_k)^*Q,\omega^A,\d H)$ is the
{\sl $k$-symplectic Hamiltonian system associated with~$L$}.
\item
If $L$ is almost-regular,
let ${\cal P}$ be the image of the Legendre map,
and $\jmath_0\colon{\cal P} \hookrightarrow (\Tan^1_k)^*Q$
the corresponding embedding,
and denote by $FL_0 \colon \Tan^1_kQ \to {\cal P}$ the restriction of
the Legendre map defined by $\jmath_0 \circ FL_0 = FL$.
Then, the condition of almost-regularity implies that
there exists $H_0\in\Cinfty({\cal P})$ such that
$(FL_0)^*(H_0) = E_L$.
Furthermore, we can define $\omega^A_0=\jmath_0^*(\omega^A)$.
With these definitions,
the triple $({\cal P},\omega^A_0,\d H_0)$ is the
{\sl $k$-presymplectic Hamiltonian system associated with~$L$},
and the corresponding Hamiltonian field equation (\ref{generic0}) is
$$
\inn(X^0_A) \omega^A_0 = dH_0
$$
where ${\bf X}^0=(X^0_1,\ldots,X^0_k)$ (if it exists)
is a $k$-vector field on ${\cal P}$.
Once again, in general, this equation has no
solutions everywhere in ${\cal P}$ but, in the most favourable situations,
they do in a submanifold of ${\cal P}$ which is obtained applying
the $k$-presymplectic constraint algorithm developed in
Section \ref{precoal}.
\end{itemize}

\section{Applications and examples}

\subsection{The Skinner--Rusk unified formalism for
$k$-symplectic field theory}

The so-called {\sl Skinner--Rusk formalism}
\cite{skinner1,skinner2}
was developed in order to give a
geometrical unified formalism for describing mechanical systems.
It incorporates all the characteristics of Lagrangian and
Hamiltonian descriptions of these systems.
This formalism has been generalized to the $k$-symplectic description of
first-order field theories in \cite{RRS}.
Next we outline the main features of this formalism.

Let us consider the direct sum $T^1_kQ\oplus (T^1_k)^*Q$
(of vector bundles over~$Q$),
with coordinates $(q^i,v^i_A,p^A_i)$, and denote by
$pr_1 \colon T^1_kQ \oplus (T^1_k)^*Q \to T^1_kQ$ and
$pr_2 \colon T^1_kQ \oplus (T^1_k)^*Q \to (T^1_k)^*Q$
the canonical projections.
In this manifold, we have some canonical structures.

First, if
$((\omega_0)_1, \ldots , (\omega_0)_k)$ is the canonical
$k$-symplectic structure on $(T^1_k)^*Q$,
its pull-back through $pr_2$
yields a $k$-presymplectic structure
$(\Omega_1, \ldots , \Omega_k)$ on
$T^1_kQ \oplus (T^1_k)^*Q$:
the 2-forms are defined by $\Omega_A = (pr_2)^*(\omega_0)_A$.

We can also define the so-called coupling function
${\cal C} \colon T^1_kQ \oplus (T^1_k)^*Q \longrightarrow \Real$
by
$$
{\cal C}({v_1}_q,\ldots ,{v_k}_q,\alpha^1_q,\ldots ,\alpha^k_q):=
\langle \alpha^A_q , {v_A}_q \rangle .
$$

Now, consider a Lagrangian $L \in \Cinfty(T^1_kQ)$.
We can define a Hamiltonian function
$\mathcal{H} \in \Cinfty(T^1_kQ \oplus (T^1_k)^*Q)$ as
$\mathcal{H} = {\cal C} - pr_1^*(L)$:
$$
\mathcal{H}({v_1}_q,\ldots,{v_k}_q,\alpha^1_q,\ldots,\alpha^k_q)
=
{\cal C}({v_1}_q,\ldots,{v_k}_q,\alpha^1_q,\ldots,\alpha^k_q)-
L({v_1}_q,\ldots,{v_k}_q) ,
$$
which in local coordinates reads
$\mathcal{H} = \alpha_i^A \, v^i_A - L(q^i,v^i_A)$.

Then $\left( T^1_kQ \oplus (T^1_k)^*Q,\Omega_A,\mathcal{H} \right)$
is a $k$-presymplectic Hamiltonian system;
where $\ds \cap_{A=1}^k\Omega_A$
is locally generated by the vector fields
$\ds\left\{ \derpar{}{v^i_A}\right\}$.
We look for the solutions of its HDW equation
which are integral sections
$\psi \colon \Real^k \to T^1_kQ \oplus (T^1_k)^*Q$
of some integrable $k$-vector field
${\bf Z}=(Z_1,\ldots,Z_k)$
on $T^1_kQ \oplus (T^1_k)^*Q$,
satisfying
\begin{equation}
\label{s3}
\imath_{Z_A} \Omega_A = d\mathcal{H} \, .
\end{equation}
This equation gives various kinds of information.
In fact, writing locally each $Z_A$ as
$$
Z_A=  (Z_A)^i \ds\frac{\partial}{\partial q^i} +
(Z_A)^i_B\ds\frac{\partial}{\partial v^i_B}+ (Z_A)^B_i
\ds\frac{\partial}{\partial p^B_i} \ ,
$$
equation (\ref{s3}) amounts to the following conditions:
$$
p^A_i = \ds \frac{\partial L}{\partial v^i_A}\circ pr_1
\, , \quad
(Z_A)^i = v^i_A
\, , \quad
(Z_B)^B_i = \ds \frac{\partial L}{\partial q^i}\circ pr_1
\, .
$$

The first group of equations are algebraic
rather than differential,
and they define a submanifold $M_L$ of
$T^1_kQ\oplus_Q(T^1_k)^*Q$ where the equation (\ref{s3}) has
solution.
These constraints can also be obtained by computing
$\ds \inn\left(\derpar{}{v^i_A}\right)\d{\cal H}$,
as noted in the discussion of the
$k$-presymplectic constraint algorithm.
Observe that the submanifold $M_L$ is just the graph
of the Legendre map $FL$ defined by the Lagrangian $L$,
and hence it is diffeomorphic to $T^1_kQ$.
We denote by $\jmath\colon M_L \to T^1_kQ\oplus_Q (T^1_k)^*Q$ the
natural embedding.

The second group of equations are a holonomy condition
which means that the $k$-velocity part of the
integral sections of the $k$-vector field ${\bf Z}$
is the lift of a section $\phi \colon \Real^k \to Q$.

The third group of equations establishes some relations
among some of the coefficients
$(Z_A)^B_i$ of the vector fields $Z_A$.

Given a solution ${\bf Z}=(Z_1,\ldots,Z_k)$ of equation~(\ref{s3}),
the vector fields $Z_A$
are tangent to the submanifold $M_L$ if, and only if,
the functions
$\Lie_{Z_A}
\left( p^B_j -\ds\frac{\partial L}{\partial v^j_B}\circ pr_1 \right)
$
vanish at the points of $M_L$, for every $A,B,j$
(the symbol $\Lie$ denotes the Lie derivative).
Taking into account the above results, this is equivalent to
\begin{equation}
\label{s8}
(Z_A)^B_j
=
\frac{\partial^2 L}{\partial v^j_B \partial q^i} v^i_A
+
\frac{\partial^2 L}{\partial v^j_B \partial v^i_C} (Z_A)^i_C
\, .
\end{equation}

In general, equations (\ref{s3}) have not a unique solution.
If $L$ is regular, taking into account the above results,
one can define local $k$-vector fields
$(Z_1 ,\ldots , Z_k)$ on a neighborhood of each point in
$M_L$ which are solutions to (\ref{s3}).
The vector field $Z_A$ may be locally given by
$$
(Z_A)^i = v^i_A
\, , \quad
(Z_A)^B_i =
\frac{1}{k} \frac{\partial L}{\partial q^i} \, \delta_A^B \, ,
$$
with $(Z_A)^i_B$ given by equation~(\ref{s8}).
Then, using a partition of the unity,
one can construct global $k$-vector fields which are
solutions to (\ref{s3}).
When the Lagrangian $L$ is singular one cannot assure
the existence of consistent solutions for equation (\ref{s3}).
Then, in the best cases, the constraint algorithm will provide a
constraint submanifold ${\cal P}_f$ where these solutions exist.

If ${\bf Z}$ is an integrable $k$-vector field solution to (\ref{s3}),
then every integral section of ${\bf Z}$ is of the form
$\psi=(\psi_L,\psi_H)$,
with
$\psi_L = pr_1 \circ \psi \colon \Real^k \to T^1_kQ$,
and as $\psi$ takes values in $M_L$ then
$\psi_H = FL \circ \psi_L$;
in fact,
$$
\psi_H(t)=(pr_2 \circ \psi)(t)=(\psi^i(t),\psi^A_i(t))
=\left(\psi^i(t), \ds\frac{\partial L}{\partial
v^i_A}\Big\vert_{\psi_L(t)}\right) =(FL \circ \psi_L)(t) \, .
$$
Furthermore, it can be proved (see \cite{RRS}) that
$\psi_L$ is the canonical lift $\phi^{(1)}$ of the projected section
$\phi= \tau_Q \circ pr_1 \circ \psi \colon \Real^k \to Q$,
which is a solution to the Euler-Lagrange field equations,
and that, if $L$ is regular,
then $\psi_H=FL\circ\psi_L$ is a solution to the Hamilton-De
Donder-Weyl field equations, where the Hamiltonian $H$
is locally given by $H \circ FL = E_L$.
In the almost-regular case, this last result also holds,
but the sections $\psi$, $\psi_L$ and $\psi_H$ take values not on
$M_L$, $T^1_kQ$ and $(T^1_k)^*Q$, but in the final constraint
submanifold ${\cal P}_f$ and on the projection submanifolds
$pr_1({\cal P}_f)\hookrightarrow T^1_kQ$ and
$pr_2({\cal P}_f)\hookrightarrow (T^1_k)^*Q$, respectively.

In this way, every constraint, differential equation, etc.\ in the
unified formalism can be translated to the Lagrangian or the
Hamiltonian formalisms by restriction to the first or the second
factors of the product bundle.
In particular, the constraint conditions
$\ds p^A_i-\frac{\partial L}{\partial v^i_A}\circ pr_1=0$
generate, by $pr_2$-projection, the primary constraints
of the Hamiltonian formalism for singular Lagrangians
(i.e., the image of the Legendre transformation,
$FL(T^1_kQ)\subset(T^1_k)^*Q$),
and they are the primary Hamiltonian constraints.

\subsection{Example~1}

Let us study a simple example, the electromagnetic field
in 2 dimensions.
The base manifold is $Q=\Real^2$, with local coordinates $(q^1, q^2)$,
and $k=2$.
The induced coordinates on $\Tan^1_2\Real^2$ are
$(q^1, q^2,v^1_1,v^1_2,v^2_1,v^2_2)$.
The electromagnetic field Lagrangian is $L=\frac{1}{2} (v^1_2 + v^2_1)^2$
(see~\cite{KS-2001}).

The canonical 2-tangent structure on $\Tan^1_2\Real^2$, $(S^1,S^2)$, is
$S^1=\frac{\partial}{\partial v^1_1} \otimes \d q^1
        +\frac{\partial}{\partial v^2_1} \otimes \d q^2$ and
$S^2=\frac{\partial}{\partial v^1_2} \otimes \d q^1
        +\frac{\partial}{\partial v^2_2} \otimes \d q^2$,
and the Liouville vector field reads as
$\Delta =
\Delta_1 +\Delta_2= v^1_1 \derpar{}{v^1_1} + v^2_1 \derpar{}{v^2_1}
 + v^1_2 \derpar{}{v^1_2} + v^2_2 \derpar{}{v^2_2}$.

The Lagrangian forms are
$$
\theta_L^1=  \d L \circ S^1 = (v^1_2 + v^2_1)\d q^2
\quad , \quad
\theta_L^2=  \d L \circ S^2 = (v^1_2 + v^2_1)\d q^1,
$$
$$
\omega_L^1=-\d\theta_L^1 =
\d q^2 \wedge \d v^1_2 + \d q^2 \wedge \d v^2_1
\quad , \quad
\omega_L^2=-\d\theta_L^2 =
\d q^1 \wedge \d v^1_2 + \d q^1 \wedge \d v^2_1,
$$
and the Lagrangian energy function is
$$
E_L=\Delta (L) -L = (v^1_2 + v^2_1)^2 - \frac{1}{2} (v^1_2 + v^2_1)^2 =
\frac{1}{2} (v^1_2 + v^2_1)^2.
$$

Since
$\Ker\omega^1_L \cap \Ker\omega^2_L =
\left\langle\derpar{}{v^1_1},\derpar{}{v^2_2}\right\rangle$,
$L$ is not regular and
$(\Tan^1_2\Real^2,(\omega_L^1,\omega_L^2),\d E_L)$ is a
$2$-presymplectic Hamiltonian system.

The field equation is
$$\inn(X_1)\omega_L^1 + \inn(X_2)\omega_L^2 =\d E_L\,$$
for a 2-vector field ${\bf X}=(X_1,X_2)$ on $\Tan^1_2\Real^2$.

If we write in coordinates
$$
X_1 =
(X_1)^1\derpar{}{q^1} + (X_1)^2\derpar{}{q^2} +
(X_1)^1_1\derpar{}{v^1_1} + (X_1)^1_2\derpar{}{v^1_2} +
(X_1)^2_1\derpar{}{v^2_1} + (X_1)^2_2\derpar{}{v^2_2},
$$
$$
X_2 =
(X_2)^1\derpar{}{q^1} + (X_2)^2\derpar{}{q^2} +
(X_2)^1_1\derpar{}{v^1_1} + (X_2)^1_2\derpar{}{v^1_2} +
(X_2)^2_1\derpar{}{v^2_1} + (X_2)^2_2\derpar{}{v^2_2},
$$
then the field equation reads as
$$
(X_1)^2 (\d v^1_2 + \d v^2_1) - ((X_1)^1_2 + (X_1)^2_1) \d q^2 +
  (X_2)^1 (\d v^1_2 + \d v^2_1) - ((X_2)^1_2 + (X_2)^2_1) \d q^1 =
  (v^1_2 + v^2_1)(\d v^1_2 + \d v^2_1).
$$
We obtain that
$$
(X_1)^2 + (X_2)^1 = v^1_2 + v^2_1,
\quad
(X_1)^1_2 + (X_1)^2_1 = 0,
\quad
(X_2)^1_2 + (X_2)^2_1 =0.
$$
Since $\d E_L \in (\Ker\omega^1_L \cap \Ker\omega^2_L)^0$,
there are no constraints and the equation has solutions
at the whole manifold $\Tan^1_2\Real^2$.
The general solution has the form
$$
X_1 =
(X_1)^1 \derpar{}{q^1} + (v_1^2+A)\derpar{}{q^2} +
B \derpar{}{v^1_1} + C \derpar{}{v^1_2} -
C \derpar{}{v^2_1} + D \derpar{}{v^2_2},
$$
$$
X_2 =
(v^1_2-A) \derpar{}{q^1} + (X_2)^2\derpar{}{q^2} +
E \derpar{}{v^1_1} + F \derpar{}{v^1_2} -
F \derpar{}{v^2_1} + G \derpar{}{v^2_2},
$$
where $(X_1)^1, (X_2)^2, A, B, C, D, E, F$ and $G$ are arbitrary functions.

In order that the integral sections of solutions be holonomic,
$$
(X_1)^1=v^1_1,\quad (X_2)^2=v^2_2,\quad
A=0, \quad C=E \;\;\;\mbox{\rm and}\;\;\; D=-F ,
$$
and, furthermore, we must also demand that
$[ X_1, X_2 ] =0$.

Now we will study the Hamiltonian formalism.
Let $(q^1, q^2, p^1_1, p^2_1, p^1_2, p^2_2)$ be the induced coordinates on
$(\Tan^1_2)^*\Real^2$.
The Legendre map
$FL\colon \Tan^1_2\Real^2 \longrightarrow (\Tan^1_2)^*\Real^2$
locally reads
$$
FL(q^1,q^2,v^1_1,v^1_2,v^2_1,v^2_2)=
(q^1,q^2,p^1_1=0,p^2_1=v^1_2+v^2_1,p^1_2=v^1_2+v^2_1,p^2_2=0).
$$

The image of $FL$,
$\mathcal{P}:= FL(\Tan^1_kQ) =\{p^1_1=0, p^2_1=0, p^2_1=p^1_2\}$,
is a submanifold of $(\Tan^1_2)^*\Real^2$.
Let $\jmath_0(q^1,q^2,p)=(q^1,q^2,0,p,p,0)$ be the natural embedding and
$FL_0\colon \Tan^1_kQ\to{\cal P}$ the restriction of the Legendre map.
We have the Hamiltonian function $H_0=\frac{1}{2} p^2$
(which is such that $(FL_0)^*H_0=E_L$),
and the 2-forms
$$
\omega^1_0=\jmath_0^*\omega^1 =
\jmath_0^*(\d q^1\wedge\d p^1_1 +\d q^2\wedge\d p^1_2) =
\d q^2\wedge\d p ,
$$
$$
\omega^2_0=\jmath_0^*\omega^2 =
\jmath_0^*(\d q^1\wedge\d p^2_1 +\d q^2\wedge\d p^2_2) =
\d q^1\wedge\d p .
$$

With these definitions, $({\cal P},\omega^1_0,\omega^2_0\,\d H_0)$
is the $2$-presymplectic Hamiltonian system associated with~$L$.
The corresponding Hamiltonian field equation is
$$\inn(Y_1)\omega_0^1 + \inn(Y_2)\omega_0^2 =\d H_0,$$
where ${\bf Y}=(Y_1,Y_2)$ is a $2$-vector field on ${\cal P}$.

If, in coordinates,
$$
Y_1 = (Y_1)^1\derpar{}{q^1} + (Y_1)^2\derpar{}{q^2} + (Y_1)^0\derpar{}{p},
$$
$$
Y_2 = (Y_2)^1\derpar{}{q^1} + (Y_2)^2\derpar{}{q^2} + (Y_2)^0\derpar{}{p},
$$
the equation is
$$
(Y_1)^2\d p - (Y_1)^0 \d q^2 + (Y_2)^1\d p - (Y_2)^0 \d q^1 =  p \d p
$$
and we obtain
$$
(Y_1)^2 + (Y_2)^1 = p ,
\quad
(Y_1)^0 = 0 ,
\quad
(Y_2)^0 =0 .
$$
Since $\Ker\omega^1_0 \cap \Ker\omega^2_0=\{0\}$,
there are no constraints and the equation has solutions
at the whole manifold ${\cal P}$.
The general solution has the form
$$
Y_1 = (Y_1)^1\derpar{}{q^1} + \left(\frac{1}{2}p + A\right)\derpar{}{q^2},
$$
$$
Y_2 = \left(\frac{1}{2}p - A\right)\derpar{}{q^1} + (Y_2)^2\derpar{}{q^2},
$$
where $(Y_1)^1, (Y_2)^2$ and $A$ are arbitrary functions on ${\cal P}$.

\subsection{Example~2}

In this example we consider two independent variables
$(t,s) \in \R^2$, thus $k=2$.
The field components (dependent variables) are
$(q,e) \in Q = \R^d \times \R^+$.
The corresponding natural coordinates of
$\oplus^2 \Tan Q$ are written
$(q,e;q_t,q_s,e_t,e_s)$,
and those of
$\oplus^2 \Tan^*Q$ are
$(q,e;p^t,p^s,\pi^t,\pi^s)$.

We consider as Lagrangian function
$$
L = \frac{1}{2e} (q_t)^2 + \frac{1}{2} m^2 e - \frac{\tau}{2} (q_s)^2 ,
$$
with $m$, $\tau$ parameters,
and for instance $(q_t)^2$ is the square of $q_t$
with respect to the Euclidean inner product of~$\Real^d$.
From $L$ we compute the Lagrangian energy
$$
E_L = \frac{1}{2e}(q_t)^2-\frac{\tau}{2}(q_s)^2-\frac{1}{2}m^2e
$$
and the Legendre map
$FL \colon \oplus^2 \Tan Q \to \oplus^2 \Tan^*Q$:
$$
FL(q,e,q_t,q_s,e_t,e_s)=
\left( q,e,\frac{1}{e}q_t,-\tau q_s,0,0 \right) \ .
$$

It is clear that
the primary Hamiltonian constraint submanifold
$\mathcal{P}_0 \subset \oplus^2 \Tan^*Q$
is described by the primary hamiltonian constraints
$$
\pi^t \weak{\mathcal{P}_0} 0 ,
\quad
\pi^s \weak{\mathcal{P}_0} 0 .
$$
This also shows that the Lagrangian $L$ is almost-regular.

\subsubsection*{Hamiltonian formalism}

Using $(q,e,p^t,p^s)$ as coordinates on the submanifold~$\mathcal{P}_0$,
its $2$-presymplectic structure
---the pull-back ot the canonical 2-symplectic structure of
$\oplus^2 \Tan^*Q$---
is given by
$\omega^t_0 = \d q \wedge\d p_q^t$
and
$\omega^s_0 = \d q \wedge\d p_q^s$
---in these expressions a summation over the invisible vector indices
of $q$ and the momenta is implicit.
Then
$$
\Ker\omega^t_0 \,\cap\, \Ker\omega^s_0 =
\left\langle \derpar{}{e} \right\rangle .
$$

The Hamiltonian function on $\mathcal{P}_0$ is
$$
H_0 =
\frac{e}{2} (p^t)^2 - \frac{1}{2} m^2 e - \frac{1}{2\tau} (p^s)^2 .
$$

Consider
${\bf X}=(X_t,X_s)$,
a $2$-vector field on $\mathcal{P}_0$:
\beann
X_t &=&
F_t\derpar{}{q}+f_t\derpar{}{e}+F_t^t\derpar{}{p^t}+F_t^s\derpar{}{p^s}
\\
X_s &=&
F_s\derpar{}{q}+f_s\derpar{}{e}+F_s^t\derpar{}{p^t}+F_s^s\derpar{}{p^s}
\eeann
(where the capital $F$'s are also vector functions).
The Hamiltonian field equation for it is
$\inn(X_t)\omega_0^t + \inn(X_s)\omega^s_0 =\d H_0$:
$$
F_t \d p^t + F_s \d p^s - (F_t^t+F^s_s) \d q =
e\,p^t \d p^t - \frac{p^s}{\tau} \d p^s +
\frac{1}{2} \left((p^t)^2-m^2\right) \d e
\,,
$$
which partly determines the coefficients of~${\bf X}$:
$$
F_t = e \,p^t \, , \
F_s = -\frac{1}{\tau} p^s \, , \
F_t^t+F^s_s = 0 \, ,
$$
and imposes as a consistency condition
the secondary hamiltonian constraint
$$
\frac12 \left( (p^t)^2 - m^2 \right) \weak{\mathcal{P}_1} 0 ;
$$
this can also be obtained as
$
\xi=\inn\left(\derpar{}{e}\right)\d H = \frac12 ((p^t)^2-m^2) \weak{} 0
$.

\smallskip
Imposing the tangency of
${\bf X}$
to
$\mathcal{P}_1$
yields no more constraints
and determines partly some coefficients of~${\bf X}$:
$$
\Lie_{X_t}\xi = p^t F_t^t \weak{} 0 ,
\quad
\Lie_{X_s}\xi = p^t F_s^t \weak{} 0 ,
$$
from which the final dynamics on $\mathcal{P}_1$ is given by
\beann
X_t &=&
e\,p^t \derpar{}{q} + f_t \derpar{}{e} +
F_t^t \derpar{}{p^t} + F_t^s \derpar{}{p^s}
\\
X_s &=&
-\frac{1}{\tau}p^s \derpar{}{q} + f_s \derpar{}{e} +
F_s^t \derpar{}{p^t} - F_t^t \derpar{}{p^s} ,
\eeann
with $f_t$, $f_s$, $F_t^s$ arbitrary functions,
and $F_t^t$, $F_s^t$ arbitrary but orthogonal to $p^t$.

Consider the particular case of $d=1$
---the $q$ variable is just a scalar.
The submanifold $\mathcal{P}_1$ is given by the constraint
$p^t = m$ (or $p^t = -m$).
Then, in coordinates $(q,e,p^s)$, the dynamics reads
\beann
X_t &=&
me \derpar{}{q} + f_t \derpar{}{e} + F_t^s \derpar{}{p^s}
\\
X_s &=&
-\frac{1}{\tau}p^s \derpar{}{q} + f_s \derpar{}{e} .
\eeann
The analysis of the integrability of the 2-vector field
${\bf X} = (X_t,X_s)$
relies on the computation of 
$$
[X_t,X_s] =
- \left( \frac{1}{\tau} F^s_t + mf_s \right) \derpar{}{q} + 
\left( \Lie_{X_t} f_s - \Lie_{X_s} f_t \right) \derpar{}{e} -
\left( \Lie_{X_s} F_t^s \right)
\derpar{}{p^s} 
\,.
$$
Setting it to zero determines
$F^s_t = -\tau m f_s$
and a set of two nonlinear PDEs for $f_s$,~$f_t$:
$$
-\frac{1}{\tau}p^s \derpar{f_s}{q} + f_s \derpar{f_s}{e} = 0 \,,
$$
$$
me \derpar{f_s}{q} + f_t \derpar{f_s}{e} - \tau m f_s \derpar{f_s}{p^s}
+ \frac{1}{\tau}p^s \derpar{f_t}{q} - f_s \derpar{f_t}{e} = 0 \,.
$$
Certainly there are solutions to this equations,
as for instance the one given by $f_t=f_s=0$.
However, it does not seem easy to give an explicit
description of the whole set of these solutions.

Finally, once one has an integrable 2-vector field~${\bf X}$,
a map $\psi \colon \R^2 \to \mathcal{P}_1$,
$(t,s) \mapsto (q,e,p^s)$,
is an integral section iff it satisfies
$$
\derpar{q}{t}   = me ,
\quad
\derpar{q}{s}   = -\frac{p^s}{\tau} ,
\quad
\derpar{e}{t}   = f_t ,
\quad
\derpar{e}{s}   = f_s ,
\quad
\derpar{p^s}{t} = F_t^s ,
\quad
\derpar{p^s}{s} =0 .
$$

\subsubsection*{Lagrangian formalism}

The Lagrangian analysis can be performed in a similar way.
Let us describe it more briefly.
Using natural coordinates $(q,e;q_t,e_t,q_s,e_s)$ on $\oplus^2 \Tan Q$,
the $2$-presymplectic structure induced by~$L$ is described by
$\ds
\omega^t =
\frac{1}{e} \, \d q \wedge \d q_t - \frac{1}{e^2} q_t \, \d q \wedge \d e$
and
$\omega^s = -\tau \, \d q \wedge \d q_s$.
Then
$$
\Ker\omega^t \,\cap\, \Ker\omega^s =
\left\langle
\derpar{}{e_t} , \derpar{}{e_t} , e\derpar{}{e} + q_t \derpar{}{q_t}
\right\rangle .
$$

An arbitrary $2$-vector field ${\bf X}=(X_t,X_s)$ on $\oplus^2 \Tan Q$
reads
\beann
X_t &=&
F_t\derpar{}{q}+f_t\derpar{}{e}+
F_{tt}\derpar{}{q_t}+F_{ts}\derpar{}{q_s}
f_{tt}\derpar{}{e_t}+f_{ts}\derpar{}{e_s} ,
\\
X_s &=&
F_s\derpar{}{q}+f_s\derpar{}{e}+
F_{st}\derpar{}{q_t}+F_{ss}\derpar{}{q_s}
f_{st}\derpar{}{e_t}+f_{ss}\derpar{}{e_s} .
\eeann
If it has to satisfy the second-order condition,
one has moreover
$$
F_t=q_t ,
\quad
F_s=q_s
\quad
f_t=e_t ,
\quad
f_s=e_s .
$$

The field equation for it is
$\inn(X_t)\omega^t + \inn(X_s)\omega^s = \d E_L$
(maybe on a certain submanifold).
This determines some of the coefficients
and defines just one primary Lagrangian constraint,
$\ds
\chi = \frac12 \left( \frac{(q_t)^2}{e^2} - m^2 \right)
$.
The tangency to this submanifold does not yield new constraints,
and some functions in ${\bf X}$ remain arbitrary.
This happens regardless of whether we impose the second-order
condition or not,
the only difference being in the number of remaining
arbitrary functions.

\subsection*{Acknowledgments}

We acknowledge the financial support of the
\emph{Ministerio de Educaci\'on y Ciencia},
projects
MTM\,2005--04947, MTM\,2008--00689/MTM and MTM\,2008--03606--E/MTM.



\end{document}